\documentclass[12pt]{iopart}

\expandafter\let\csname equation*\endcsname\relax 
\expandafter\let\csname endequation*\endcsname\relax 

\usepackage{mathtools}
\usepackage{color}
\usepackage{graphicx}

\newcommand{\ip}[1]{\langle #1 \rangle}

\newcommand{\la}{\lambda}
\newcommand{\lat}{\widetilde{\lambda}}
\newcommand{\de}[2]{\delta^{#1}(#2)}
\newcommand{\bra}[1]{\langle #1|}
\newcommand{\ket}[1]{|#1\rangle}

\begin{document}
	
	\title[A connection between $\mathcal{R}$-invariants and $R$-operators in $\mathcal{N}=4$ sYM]{A connection between $\mathcal{R}$-invariants and Yang-Baxter $R$-operators in $\mathcal{N}=4$ super-Yang-Mills theory}
	
	\author{V. K. Kozin$^{1,2,3}$}
	
	
	\address{
		$^1$Science Institute,
		University of Iceland, Dunhagi 3, IS-107, Reykjavik, Iceland\\
		$^2$ITMO University, Kronverkskiy prospekt 49, Saint Petersburg 197101, Russia \\
		$^3$St. Petersburg Academic University of the Russian Academy of Sciences,
		194021 St. Petersburg, Russia\\
	}
	
	\ead{vak13@hi.is}
	
	\begin{abstract}
		The BCFW recursion relation in $\mathcal{N}=4$ super-Yang-Mills theory is solved using Yang-Baxter $R$-operators in the NMHV sector. Explicit expressions for $\mathcal{R}$-invariants are obtained in terms of the chains of $R$-operators acting on an appropriate basic state.
	\end{abstract}
	
	\section{Introduction}
	The $R$-operator formalism introduced in \cite{derk} establishes a connection between calculating the scattering amplitudes in $\mathcal{N}=4$ super-Yang-Mills theory and integrable systems, such as spin chains. This approach exploits Yangian symmetry of the amplitudes, that has been studied e.g. in~\cite{KORCHEMSKY2010377},\cite{Drummond2010} The framework of $R$-operators was developed in a number of papers \cite {ref1}, \cite{ref2}, \cite{ref3}, \cite{Bork2017}. For example, in \cite{ref3}, a connection was established between the graded permutations encoding the on-shell graphs and chains of $R$-operators acting on a suitable \textit{basic state}, as well as a connection to the top-cell graphs.
	
	As it has been shown in~\cite{derk} the amplitude terms can be obtained by acting on \textit{basic states} (formed by products of delta-functions) by products of Yang-Baxter $R$-operators. These operators are defined from the $L$-matrices by the $RLL$-intertwining relation. The $R$-operators act  on just one pair of the spin chain sites. The sequential action by Yang-Baxter R-operators on the basic state results in a production of non-local, entangled solutions, reproducing the amplitude.
	
	In order to continue the program of studying $\mathcal{N}=4$ sYM by the methods used for integrable models in quantum theory, this paper aims to develop further the formalism of $R$-operators for constructing the tree amplitudes in $\mathcal{N}=4$ sYM in the NMHV sector and get an expression for $\mathcal{R}$-invariants through $R$-operators.
	
	The paper is organized as follows. In Section~\ref{sec:amp} we introduce the basic notation for the scattering amplitudes in $\mathcal{N}=4$ sYM. In Section~\ref{sec:spin_chain} we show the connection according to~\cite{derk} between the $gl(4|4)$ spin chains and scattering amplitudes in $\mathcal{N}=4$ sYM, and introduce the notation for the main objects of the paper -- Yang-Baxter $R$-operators (which will be referred to as $R$-operators) and discuss their properties. In Section~\ref{sec:solving_BCFW} we provide a solution for the BCFW~\cite{ PhysRevLett.94.181602} recursive relation in $\mathcal{N}=4$ sYM in the NMHV sector in terms of $R$-operators and give the formulas $\mathcal{R}$-invariants, presenting them as the chains of $R$-operators acting on an appropriate \textit{basic state}, which is the main result of the paper. 
	
	\section{Amplitudes in $\mathcal{N}=4$ sYM}
	\label{sec:amp}
	The fact that the theory $\mathcal{N}=4$ sYM is supersymmetric allows one to introduce a superfield that combines all the fields into one function defined on the on-shell superspace~\cite{scat} $(\la^{\alpha},\lat_{\dot{\alpha}},\eta^A)$
	\begin{equation}
	\Phi(\la,\lat,\eta)=g^{+}+\eta^A\psi_A+\frac{1}{2!}\eta^A\eta^B\phi_{AB}+\frac{1}{3!}\epsilon_{ABCD}\eta^A\eta^B\eta^C\overline{\psi}^D+\frac{1}{4!}\epsilon_{ABCD}\eta^A\eta^B\eta^C\eta^D g^{-} 
	\end{equation}
	where the capital Latin letters $A, B, C, D$ denote the indices of the fundamental representation of the group $SU(4)_R$, and $\epsilon_{ABCD}$ is the Levi-Civita symbol, $\eta^A$ are Grassmann variables. With the help of superfields it is possible to construct a superamplitude --- a generating function for all possible scattering amplitudes of a given order:
	\begin{align}
	&M_n(\Phi_1,\dots,\Phi_n) \equiv M_n((\la_1,\lat_1,\eta_1),\dots,(\la_n,\lat_n,\eta_n))\equiv\nonumber\\
	&\equiv M_n((p_1,\eta_1),\dots,(p_n,\eta_n))\equiv M_n(1,\dots,n)
	\end{align}
	It can be shown, according to \cite{scat}, that the general form for the scattering amplitude of $n$ particles in $\mathcal{N}=4$ sYM is
	\begin{equation}
	M_n(\{\la_i,\lat_i,\eta_i\})=\frac{\delta^4(p)\delta^8(q)}{\langle12\rangle\langle23\rangle...\langle n1\rangle}P_n(\{\la_i,\lat_i,\eta_i\})
	\end{equation}
	where $p=p_1+\ldots+p_n$ -- total momentum, $q= q_1+\ldots+q_n=|1\rangle\eta_1+\ldots+|n\rangle \eta_n$ -- total supermomentum. 	The spinors $\la_{\alpha}:=\ket{p}$ and $\lat^{\dot{\alpha}}:=|p]$ correspond to the states with helicity $\pm1/2$ respectively. $P_n(\{\la_i,\lat_i,\eta_i\})$  has the form of a polynomial in $\eta_i$ and allows to classify superamplitudes (will be referred to as amplitudes hereinafter) by the type $\text{N}^{k-2}$MHV
	\begin{equation}
	\begin{array}{cccccccc}
	P_n(\{\la_i,\lat_i,\eta_i\})=&P_n^{(0)}&+&P_n^{(4)}&+&P_n^{(8)}&+\dots+&P_n^{(4n-16)}\\
	&\downarrow & & \downarrow & & \downarrow & & \downarrow\\
	&\text{MHV} & & \text{NMHV} & & \text{N}^2\text{MHV} & & \overline{\text{MHV}}
	\end{array}
	\end{equation}
	$P_n^{(0)}=1$ and $P_n^{(l)}\sim \mathcal{O}(\eta^l)$. 
	This, in particular, implies the Park-Taylor formula \cite{park} for MHV amplitudes in $\mathcal{N}=4$ sYM 
	\begin{equation}\label{eq:park}
	M_{2,n}^{\text{MHV}}=\frac{\delta^4(p)\delta^8(q)}{\langle12\rangle\langle23\rangle...\langle n1\rangle}
	\end{equation}
	where $M_{k,n}$ denotes $\text{N}^{k-2}$MHV scattering amplitude of $n$ particles, i.e. $M_{2,n}$ corresponds to MHV, $M_{3,n}$ -- NMHV etc. The amplitude type is sometimes additionally indicated as above, for example $M^{\text{MHV}}_{2,n}$.

	The introduction of a superamplitude allows one to introduce an analogue of the BCFW relations in $\mathcal{N}=4$ sYM, the so-called super-BCFW relation~\cite{scat}.
	The analytic form of the super-BCFW for $\text{N}^{k-2}$MHV amplitude is	
	\begin{align}
	M_{k;n}(1,2,...,n)&=\sum_{\substack{n_L+n_R=n+2\\k_L+k_R=k+1}}\int d^4 P 
	d^4 \eta M_L((\hat{p}_1,\hat{\eta}_1),\dots,(p_{n_L-1},\eta_{n_L-1}),(p,\eta))\frac{1}{P^2}\cdot\nonumber\\  
	&\cdot M_R((-p,\eta),(p_{n_R+1},\eta_{n_R+1}),\dots,(\hat{p}_n,\hat{\eta}_n)) 
	\end{align}
	where $p=P+z_{P_L}\la_1\lat_n$ and $z_{P_L}=\frac{P_L^2}{\bra{1}P_L|n]}$, whereas the subamplitudes $M_L$ and $M_R$ include momentum $\delta$-functions. 
	\subsection{$\mathcal{R}$-invariants and dual superconformal symmetry}
    The super-BCFW recursive relation can be solved in general for tree amplitudes. The general analytic expression for tree NMHV amplitudes in $\mathcal{N}=4$ sYM was initially obtained in the paper~\cite{DRUMMOND2010317}
	\begin{equation}\label{eq:nmhv}
	M^{\text{NMHV}}_{3,n}=M^{\text{MHV}}_{2,n}\sum_{\substack{1<s<t<n\\ |s-t|\ge 2}}R_{n;st}
	\end{equation}
	where $\mathcal{R}_{r;st}$ -- dual superconformal invariants ($\mathcal{R}$-invariants). The explicit form of $\mathcal{R}_{r;st}$ is
	\begin{equation}\label{eq:rinv}
	\mathcal{R}_{r;st} = \frac{\ip{s s-1}\ip{t t-1}\de{4}{\Xi_{r;st}}}{x_{st}^2\bra{r}x_{rs}x_{st}\ket{t}
		\bra{r}x_{rs}x_{st}\ket{t-1}
		\bra{r}x_{rt}x_{ts}\ket{s}
		\bra{r}x_{rt}x_{ts}\ket{s-1}
	}
	\end{equation}
	where $x_{ab}:=p_a+\ldots+p_{b-1}$, $\theta_{ab}:=q_a+\ldots+q_{b-1}$ are dual variables. At $b<a$ we have $x_{ab}=-x_{ba}$.
	The Grassmann-odd quantity $\Xi_{r;st}$ is defined by
	$$
	\Xi_{r;st}:=\bra{r}x_{rs}x_{st}\ket{\theta_{tr}}+\bra{r}x_{rt}x_{ts}\ket{\theta_{sr}}
	$$
	Expressions of the form $\bra{r}x_{rs}x_{st}\ket{t}$ should be interpreted as  $\bra{r}^a (x_{rs})_{a\dot{c}}(x_{st})^{\dot{c}b}\ket{t}_b$.
	
	In the paper~\cite{yangian} it has been shown that it is possible to combine the algebras of superconformal and dual superconformal symmetries of tree scattering amplitudes in $\mathcal{N}=4$ sYM into an infinite-dimensional algebra called Yangian $Y(psu(2,2|4))$. Then the tree amplitudes will be the sum of the Yangian invariants in the super-BCFW decomposition (which will be referred to hereafter as BCFW).
	\section{Spin chains and on-shell graphs}
	\label{sec:spin_chain}
	In the paper~\cite{derk}, each tree scattering amplitude of $n$ particles $M_n$ in $\mathcal{N}=4$ sYM is associated with a $gl(4|4)$ spin chain of length $n$. As we know from the paper~\cite{fadd}, a discrete set of canonically conjugate coordinates and momenta can be associated with a discrete set of spins, thus forming a spin chain. The paper~\cite{derk} introduces a set of canonical variables
	$\boldsymbol{x}=(x_a)^{N+M}_{a=1}$, $\boldsymbol{p}=(p_a)^{N+M}_{a=1}$, satisfying the commutation relations $\{x_a,p_b]=-\delta_{ab}$ where $\{x_a,p_b]$ is a graded commutator, $N$ and $M$ are correspondingly the numbers of bosonic and fermionic components.
	
	The spin chain is an example of an integrable quantum  model, and one can apply quantum inverse scattering method (QISM) (see, for example, papers by L. D. Faddeev and collaborators~\cite{fadd}, \cite{Fadd2}, \cite{Kulish:1981bi}, \cite{Kulish1981}) to solve it. One of the central objects of QISM is the monodromy matrix
	\begin{equation}
	[T(u)]_{ac}=[L_1(u)]_{ab_1}[L_2(u)]_{b_1b_2}\dots[L(u)]_{b_{n-1}c}
	\end{equation}
	where $L$-operators 
	\begin{equation}
	[L(u)]_{ab}=u\delta_{ab}+x_a p_b.
	\end{equation}	
	Further, the authors of the paper~\cite{derk} introduce the $R$-operators defined by the RLL-relation	
	\begin{equation}
	R_{12}(u-v)[L_1(u)]_{ab}[L_2(v)]_{bc}=
	[L_1(v)]_{ab}[L_2(u)]_{bc}R_{12}(u-v)
	\end{equation}
	and give the solution to the RLL-relation for $gl(4|4)$ (relevant to $\mathcal{N}=4$ sYM)
	\begin{equation}\label{eq:r_op}
	R_{12}(u)=\int_{0}^{+\infty}\frac{dz}{z^{1-u}}e^{-z(\boldsymbol{p_1}\cdot \boldsymbol{x_2})}
	\end{equation}
	Finally, the connection of a spin chain with $\mathcal{N}=4$ sYM is established by the following definition of canonically conjugate variables
	\begin{align}
	\boldsymbol{x}:=(\lambda_{\alpha}, \partial_{\widetilde{\lambda}_{\dot{\alpha}}}, \partial_{\eta_A})\\
	\boldsymbol{p}:=(\partial_{\lambda_{\alpha}}, -\widetilde{\lambda}_{\dot{\alpha}}, -\eta_A)
	\end{align}
	Then the action of the operator $R_{ij}(u)$ on an arbitrary function $F(\la_i,\lat_i,\eta_i,\la_j,\lat_j,\eta_j)$ is given by~\cite{derk}
	\begin{equation}
	R_{ij}(u)F(\la_i,\lat_i,\eta_i,\la_j,\lat_j,\eta_j)=
	\int_0^{+\infty}\frac{dz}{z^{1-u}}F(\la_i-z\la_j,\lat_i,\eta_i,\la_j,\lat_j+z\lat_i,\eta_j+z\eta_i)
	\end{equation}
	that is, it performs a BCFW shift on the spinor-helicity variables. Also, as shown in~\cite{derk}, the condition of Yangian invariance of scattering amplitudes in $\mathcal{N}=4$ sYM is formulated in that way, that the amplitude $M$ is an eigenfunction of the monodromy operator
	\begin{equation}
	T(u) M=C \cdot M
	\end{equation}
	where the eigenvalue $C$ plays a minor role. Thus, a connection is established between the scattering amplitudes in $\mathcal{N}=4$ sYM and $gl(4|4)$ spin chains.
	
	The $R$-operator from the equation~(\ref{eq:r_op}) will be the main construction object of amplitudes, allowing us to construct scattering amplitudes in the spirit of the QISM method. The essentially non-local object --- amplitude $M_n$, depending on the variables of all $n$ external particles will be built using the product of $R$ -operators, each of which acts only on a pair of variables associated with external particles. In~\cite {derk}, the authors give the formulas for 3-particle scattering amplitudes $M^{\text{MHV}}_{2,3}$ and $M^{\overline{\text{MHV}}}_{1,3}$ in $\mathcal{N}=4$ sYM, expressed in $R$-operators as 
	\begin{equation}
	\begin{aligned}
	M^{\overline{\text{MHV}}}_{1,3}=R_{12}R_{23}\Omega_{1,3}\quad\mathrm{and}\quad\Omega_{1,3}=\delta^2(\la_1)\delta^2(\la_2)\delta^2(\lat_3)\delta^4(\eta_3),
	\end{aligned}
	\end{equation}
	\begin{equation}
	\begin{aligned}
	M^{\text{MHV}}_{2,3}=R_{23}R_{12}\Omega_{2,3}\quad\mathrm{and}\quad\Omega_{2,3}=\delta^2(\la_1)\delta^2(\lat_2)\delta^4(\eta_2)\delta^2(\lat_3)\delta^4(\eta_3),
	\end{aligned}
	\end{equation}
	where $R_{ij}\equiv R_{ij}(0)$.
	
	\subsection{On-shell diagrams}   
\begin{figure}[ht]
		\includegraphics[width=\linewidth]{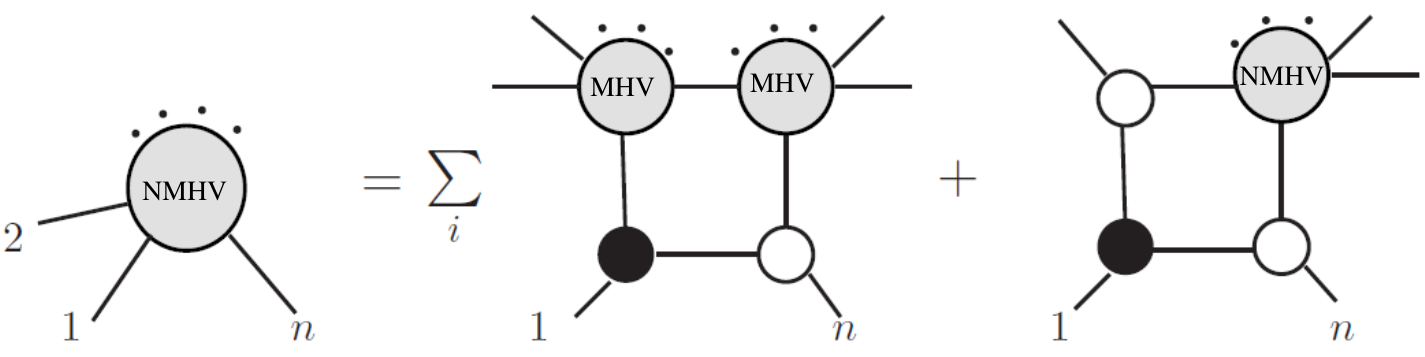}
		\caption{The BCFW recursion relation in terms of on-shell graphs for an NMHV amplitude. The summation is performed over all possible MHV subamplitudes, where the index $i$ denotes the rightmost external leg of the left MHV subamplitude.}
		\label{gr:bcfw_onshell}
	\end{figure}

	The work~\cite{nima} shows that BCFW decomposition can be written in terms of the so-called on-shell graphs. The building blocks of on-shell diagrams are 3-particle MHV and anti-MHV amplitudes depicted in Fig.~\ref{gr:bcfw_onshell} with black and white circles respectively.	
	In terms of the BCFW on-shell diagrams, the decomposition, according to~\cite{nima}, can be written diagrammatically as in Fig.~\ref{gr:bcfw_onshell}.
	where the sum is performed for all possible MHV subamplitudes (non-recurrent terms) and the last term contains the NMHV subamplitude (recurrent term).
	The right-hand side of the diagram decomposition in Fig.~\ref{gr:bcfw_onshell} looks exactly like a BCFW-diagram with added "bridges" (the so-called BCFW-bridge).
	According to~\cite{derk}, the $R_{1n}$-operator implements this bridge.
	For on-shell diagrams, the rules of diagram technique change as compared to BCFW diagrams -- here each internal line is assigned an integral $\int d^4\eta d^4 P\delta(P^2)$.
	
	\section{Solving BCFW in the NMHV sector}
	\label{sec:solving_BCFW}
\begin{figure}[ht]
		\includegraphics[width=\linewidth]{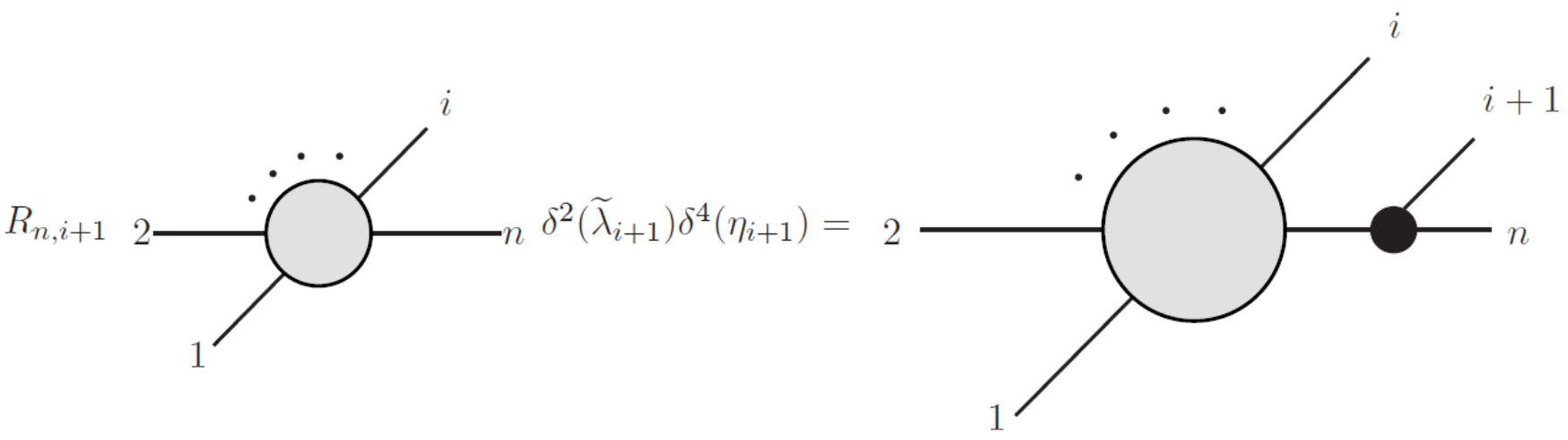}
		\caption{A diagrammatic representation of Eq.~(\ref{eq:th_alg}).}
		\label{gr:th_gr}
	\end{figure}

	To start solving the BCFW relations with the help of $R$-operators, we formulate the following statement
	\begin{align}\label{eq:th_alg}
	&R_{n,i+1}M(1,2,...,i,n)\de{2}{\lat_{i+1}}\de{4}{\eta_{i+1}}=\\ 
	&=\int d\eta_0 d^4 P_0\delta(P_0^2)M(1,2,...,i,\{\ket{-P_0},|-P_0],\eta_0\})M^{\text{MHV}}_{2,3}(\{\ket{P_0},|P_0],\eta_0\},i+1,n),\nonumber
	\end{align}
	which we will be using further. Graphically, this statement is shown in Fig.~\ref{gr:th_gr}.
	The proof of Eq.~(\ref{eq:th_alg}) is given in Appendix.
	\subsection{Non-recurrent terms of BCFW}
	Now we can proceed to the calculation of diagrams, which are non-recurrent terms in the diagram expansion of the NMHV amplitude (Fig.~\ref{gr:bcfw_onshell}). To do this, first build the amplitude $M^{\text{MHV}}_{2,t}(1,2...t-2,t-1,n)$. 
	
	We start the construction with a 3-part amplitude $M^{\text{MHV}}_{2,3}(1,t-1,n)$ and add the ends $t-2,t-3,\dots,2$ to the left using the Inverse Soft Limit (ISL) using $R$-operators~\cite{Arkani-Hamed2010, derk}. As a result, we get the amplitude $M_{2,t}(1,2,\ldots,t-2,t-1,n)$. We construct a chain of $R$-operators corresponding to the described procedure. The 3-particle amplitude $M_{2,3}(1,t-1,n)$, expressed in terms of $R$-operators, according to~\cite{derk}, is given by 
	\begin{equation} M_{2,3}(1,t-1,n)=R_{1t-1}R_{1n}\de{2}{\la_1}\de{2}{\lat_n}\de{4}{\eta_n}\de{2}{\la_{t-1}} 
	\end{equation} 
	Then, adding the particles $\{t-2,t-3,\dots,2\}$ using the ISL, we obtain the expression for $M_{2,t}(1,2,\ldots,t-2,t-1,n)$ 
	\begin{equation} M_{2,t}(1,2,\ldots,t-2,t-1,n)=R_{21}R_{23}\cdot...\cdot R_{t-21}R_{t-2t-1}R_{1t-1}R_{1n}\Omega^{t-1,n}_{1,\ldots,t-2,t-1,n} 
	\end{equation} 
	where $\Omega^{t-1,n}_{1,\ldots,t-2,t-1,n}\equiv\de{2}{\la_1}\de{2}{\la_2}\cdot...\cdot \de{2}{\la_{t-2}}\de{2}{\lat_{t-1}}\de{4}{\eta_{t-1}}\de{2}{\lat_{n}}\de{4}{\eta_n}$, and the superscripts $t-1,n$ distinguish the delta-functions containing ($\lat$, $\eta$) variables. Now, according to the proved formula~\ref{eq:th_alg}, we attach to the obtained amplitude $M_{2,t}(1,2,\ldots,t-2,t-1,n)$ the 3-particle MHV subamplitude at the outer end $n$. In the language of $R$-operators, the given transformation of the amplitude $M_{2,t}(1,2,\ldots,t-2,t-1,n)$ corresponds to the expression $R_{nt}M_{2,t}(1,2,\ldots,t-2,t-1,n)\de{2}{\lat_t}\de{4}{\eta_t}$. Further, applying the ISL, we add to the obtained on-shell diagram the external ends $t+1,t+2,\ldots,n$, which yields the following  expression
	\begin{equation} R_{n-1n-2}R_{n-1n}\cdot...\cdot R_{t+1t}R_{t+1n}\cdot R_{nt}\cdot R_{21}R_{23}\cdot...\cdot R_{t-2 1}R_{t-2t-1}R_{1t-1}R_{1n}\Omega^{t-1,t,n}_{1,\ldots,n} \end{equation} 
	where the appropriate basic state $\Omega^{t-1,t,n}_{1,\ldots,n}$ is determined 
	\begin{align} 
	&\Omega^{t-1,t,n}_{1,\ldots,n}=\\
	&\de{2}{\la_1}\de{2}{\la_2}\cdot...\cdot \de{2}{\la_{t-2}}\de{2}{\lat_{t-1}}\de{4}{\eta_{t-1}}\de{2}{\lat_{t}}\de{4}{\eta_{t}}\de{2}{\la_{t+1}}\cdot...\cdot\de{2}{\la_{n-1}}\de{2}{\lat_{n}}\de{4}{\eta_n}\nonumber 
	\end{align} 
	It remains only to turn it into a BCFW diagram by adding the BCFW bridge to the outer ends $1$ and $n$. According to~\cite{derk} such a bridge is implemented using the $R_{1n}$ operator. Thus, acting on the obtained on-shell diagram with the operator $R_{1n}$, we get the desired BCFW-diagram. This BCFW diagram, according to the paper~\cite{drum}, corresponds to the expression $M^{\text{MHV}}_{2,n}\mathcal{R}_{n;2,t}$. That is, we get the expression for $\mathcal{R}_{n;2,t}$ in $R$-operators: \begin{align} \label{eq:rchain2} 
	&M^{\text{MHV}}_{2,n}\mathcal{R}_{n;2,t}=\\
	&R_{1n}\cdot R_{n-1n-2}R_{n-1n}\cdot...\cdot R_{t+1t}R_{t+1n}\cdot R_{nt}\cdot R_{21}R_{23}\cdot...\cdot R_{t-2 1}R_{t-2t-1}R_{1t-1}R_{1n}\Omega^{t-1,t,n}_{1,2,...,n}\nonumber 
	\end{align} 
	\subsection{Recurrent term of the BCFW decomposition} 
\begin{figure}[ht] \includegraphics[width=0.6\linewidth]{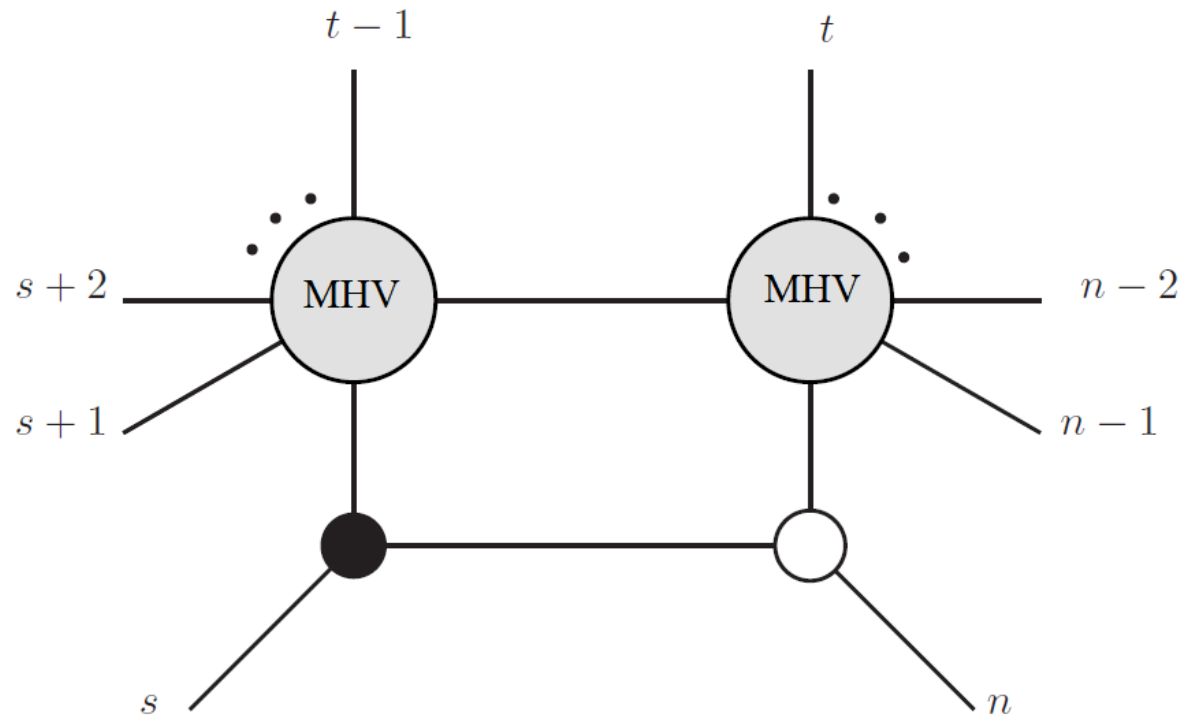} 
		\caption{Diagram arising at the second ($s=2$) recursion step (the missing ends $1...s-1$ are added with the ISL).} 
		\label{gr:diag_rec} 
	\end{figure}

	Now we turn to the last diagram in the BCFW decomposition in Fig.~\ref{gr:bcfw_onshell}. This diagram can be viewed as adding a particle with the number $1$ through the ISL to the subamplitude $M^{\text{NMHV}}_{3,n-1}(2,3,\ldots,n)$. According to~\cite{derk} this type of the ISL is realized through a pair of $R$-operators in the following form \begin{equation} \label{eq:isl2} R_{1n}R_{12}M^{\text{NMHV}}_{3,n-1}(2,3,\ldots,n)\de{2}{\la_1} 
	\end{equation} 
	The amplitude $M^{\text{NMHV}}_{3,n-1}(2,3,\ldots,n)$ in its turn, is decomposed recursively with the help of BCFW into the sum of amplitudes of the form given in Fig.~\ref{gr:diag_rec}
	and the term containing an NMHV subamplitude. The amplitude given above in Fig.~\ref{gr:diag_rec} differs from the previously calculated ones (non-recurrent terms) by re-designation of the ends, and thus corresponds to the expression 
	\begin{equation} 
	M^{\text{MHV}}_{2,n-1}(2,3,\ldots,n)\mathcal{R}_{n;3,t}.
	\end{equation} 
	Acting on it with a pair of $R_{1n}R_{12}$ operators, according to the formula~(\ref{eq:isl2}), we obtain 
	\begin{equation} R_{1n}R_{12}M^{\text{MHV}}_{2,n-1}(2,3,\ldots,n)\mathcal{R}_{n;3,t}\de{2}{\la_1}= M^{\text{MHV}}_{2,n}(1,2,\ldots,n)\mathcal{R}_{n;3,t} \end{equation} 
	because $\mathcal{R}_{n;s,t}$ does not explicitly depend on $\lat_{2}$ and $\lat_n$ , which directly follows from the expression for $\mathcal{R}$-invariants (Eq.~(\ref{eq:rinv})). Thus, with a recursion depth equal to $s$ (the first decomposition step corresponds to $s=1$), the $\mathcal{R}$ invariant of $\mathcal{R}_{n;s+1,t}$ is obtained. We construct its expression in terms of $R$-operators in the same way as was done in the first iteration of the BCFW decomposition. As a result, we obtain a general expression for an arbitrary $\mathcal{R}$-invariant $\mathcal{R}_{n;s+1,t}$ terms of $R$-operators 
	\begin{align}\label{eq:main}
	&M_{2,n}^{\text{MHV}}(1,2,...,n)\mathcal{R}_{n;s+1,t}=\nonumber\\
	&I^{\prime}_1\cdot...\cdot I^{\prime}_{s-1}\cdot R_{sn}\cdot I_{n-1}\cdot...\cdot I_{t+1}\cdot R_{nt}\cdot I^{(s)}_{s+1}\cdot...\cdot I^{(s)}_{t-2}\cdot R_{st-1}R_{sn}\Omega^{t-1,t,n}_{1,2,...,n}
	\end{align} 
	where $I^{\prime}_k\equiv R_{kn}R_{kk+1}$, $I_{k}\equiv R_{kk-1}R_{kn}$ and $I^{(s)}_{k}\equiv R_{ks}R_{kk+1}$ - for brevity, we denote the pairs of $R$-operators implementing the addition of external ends through the ISL. 
	\begin{align} 
	&\Omega^{t-1,t,n}_{1,\ldots,n}=\\
	&\de{2}{\la_1}\de{2}{\la_2}\cdot...\cdot \de{2}{\la_{t-2}}\de{2}{\lat_{t-1}}\de{4}{\eta_{t-1}}\de{2}{\lat_{t}}\de{4}{\eta_{t}}\de{2}{\la_{t+1}}\cdot...\cdot\nonumber
	\cdot\de{2}{\la_{n-1}}\de{2}{\lat_{n}}\de{4}{\eta_n}\nonumber 
	\end{align} 
	is an appropriate basic state for the chain of $R$-operators in the general formula~(\ref{eq:main}), which is the main result of the paper.
	
	\section {Discussion and conclusion} 
    The formula~(\ref{eq:main}) can be understood following way. One starts with the action of operators $R_{st-1}R_{sn}$ which generates a 3-particle MHV amplitude $M_{2,n}^{\text{MHV}}(s, t-1, n)$. Then, the chain of operators $I^{(s)}_{s+1}\cdot...\cdot I^{(s)}_{t-2}$ adds the legs $s+1,...,t-2$ resulting in an MHV amplitude $M_{2,n}^{\text{MHV}}(s,s+1,...,t-1)$. Having done that, the operator $R_{nt}$ in accordance with Eq.~(\ref{eq:th_alg}) adds an MHV subamplitude on the leg $n$ with the outer end $t$. Then, the sequence $I_{n-1}\cdot...\cdot I_{t+1}$ appends the legs $t+1,...,n-1$ and  
	$R_{sn}$ realizes the BCFW-bridge for the ends $s$ and $n$, thus we arrive at the diagram depicted in Fig.~\ref{gr:diag_rec}. Finally, the series of operators
	$I^{\prime}_1\cdot...\cdot I^{\prime}_{s-1}$ appends the legs $1,..,s-1$, finishing the construction of the general term $M_{2,n}^{\text{MHV}}(1,2,...,n)\mathcal{R}_{n;s+1,t}$ of the super-BCFW expansion in the NMHV sector. 
	
	Note that the procedure described in the paper for constructing tree amplitudes by "building up" one of the subamplitudes at the outer end of the other (merging the subamplitudes and adding the BCFW-bridge) is suitable for any tree amplitude in $\mathcal{N}=4$ sYM and not just for merging MHV amplitudes, i.e. we can express generalized $\mathcal{R}$-invariants of any order (i.e. those, that appear in the $\text{N}^k$MHV sector) through $R$-operators. 
	
	Thus, in this work we have solved the BCFW relation in the NMHV sector using $R$-operators and obtained the general expression~(\ref{eq:main}) for an arbitrary $\mathcal{R}$-invariant through the chain of $R$-operators. We see the construction of a closed formula for the generalized $\mathcal{R}$-invariants in terms of $R$-operators as the next step to study. This would give us a complete solution to the problem of finding an arbitrary tree scattering amplitude in $\mathcal{N}=4$ sYM in terms of $R$-operators in the spirit of the QISM method. 
	\section{Acknowledgements}
	V. K. Kozin thanks NORDITA for hospitality. The author thanks I. E. Shenderovich for formulating the problem and S. E. Derkachev for the fruitful discussions. 
	\appendix
	\section*{Appendix: Proof of Main Lemma}\label{appendix}
    	To begin the proof of Eq.~(\ref{eq:th_alg}), we calculate the LHS 
	\begin{equation}\label{eq:int2}
	\begin{aligned}
	&R_{n,i+1}M(1,2...i,n)\de{2}{\lat_{i+1}}\de{4}{\eta_{i+1}}=\\
	&=\int_{0}^{+\infty}\frac{dz}{z}M(1,2...i,\la_n-z\la_{i+1},\lat_n,\eta_n)\de{2}{\lat_{i+1}+z\la_n}\de{4}{\eta_{i+1}+z\eta_n}=\\
	&=\int_{0}^{+\infty}\frac{dz}{z}M\cdot\delta(z+\frac{[i+11]}{[n1]})\de{4}{\eta_{i+1}+z\eta_n}=\\
	&=\frac{[1n]}{[i+11]}\de{4}{\eta_{i+1}-\frac{[i+11]}{[n1]}\eta_n}
	M(1,2,...i,\lat_n+\frac{[i+11]}{[n1]}\lat_{i+1},\lat_n,\eta_n)\delta([i+1n])
	\end{aligned}	
	\end{equation}
	We now turn to the calculation of the right-hand side of Eq.~(\ref{eq:th_alg}). It corresponds to the algebraic expression written in the right part of the statement:
	\begin{equation}\label{eq:int1}
	\begin{aligned}
	&\int d\eta_0 d^4 P_0\delta(P_0^2)M(1,2...i,\ket{-P_0},|-P_0],\eta_0)M^{\text{MHV}}_{2,3}(\ket{P_0},|P_0],\eta_0,i+1,n)=\\
	&=\int d\eta_0 d^4 P_0\delta(P_0^2)M(1,2...i,-P_0,\eta_0)\cdot\\
	&\cdot\frac{\de{4}{P_0+P_{i+1}+P_n}\de{8}{\ket{P_0}\eta_0+\ket{i+1}\eta_{i+1}+\ket{n}\eta_n}}{\ip{P_0i+1}\ip{i+1n}\ip{nP_0}}=\\
	&=\int d\eta_0 d^4 P_0\delta(P_0^2)M(1,2...i,-P_0,\eta_0)\frac{\de{4}{P_0+P_{i+1}+P_n}}{\ip{P_0i+1}\ip{i+1n}\ip{nP_0}}\cdot\\
	&\cdot\ip{P_0i+1}^4\de{4}{\eta_0-\frac{\ip{i+1n}}{\ip{P_0i+1}}\eta_n}\de{4}{\eta_{i+1}-\frac{\ip{nP_0}}{\ip{P_0i+1}}\eta_n}=\\
	&=\int d^4 P_0 \delta(P_0^2)M(1,2...\frac{\ip{i+1n}}{\ip{P_0i+1}}\eta_n,-P_0)\frac{\ip{P_0i+1}^3}{\ip{i+1n}\ip{nP_0}}\cdot\\
	&\cdot\de{4}{\eta_{i+1}-\frac{\ip{nP_0}}{\ip{P_0i+1}}\eta_n}\de{4}{P_0+P_{i+1}+P_n}
	\end{aligned}
	\end{equation}
	From the delta-function $\de{4}{P_0+P_{i+1}+P_n}$ it follows that $-P_0=P_{i+1}+P_{n}$
	\begin{equation}\label{eq:MC1}
	-\ket{P_0}\bra{P_0}=\ket{i+1}\bra{i+1}+\ket{n}\bra{n}
	\end{equation}
	The 3-particle special kinematics~\cite{elvang} yields $[P_0|\sim[i+1|\sim[n|$ and thus
	\begin{equation}\label{eq:MC2}
	\begin{aligned}
	[i+1|\sim[n|\Rightarrow[i+1|=\frac{[i+11]}{[n1]}[n|\Rightarrow\\
	-\ket{P_0}\bra{P_0}=(\ket{n}+\frac{[i+11]}{[n1]}\ket{i+1})\bra{n}
	\end{aligned}
	\end{equation}
	Using the analytical continuation of Weyl spinors
	$|-P_0]=-|P_0]$ and $\ket{-P_0}=\ket{P_0}$ one may rewrite $-\ket{P_0}\bra{P_0}$ as $\ket{-P_0}\bra{-P_0}$. Since $[P_0|\sim[n|$ than from little group scaling~\cite{elvang} it follows, that one may assume in Eq.~(\ref{eq:MC2}) $-|P_0]=|-P_0]=|n]$ and $\ket{P_0}=\ket{-P_0}=\ket{n}+\frac{[i+11]}{[n1]}\ket{i+1}$.
	
	Integration over the variable $P_0$ is very simple, since it enters the integrand through the delta function, which imposes the restriction~(\ref{eq:MC1}). Therefore, we exclude $P_0$ everywhere in the integral. Let us start with the expression $\frac{\ip{nP_0}}{\ip{P_0i+1}}$. For this purpose we multiply~(\ref{eq:MC1}) from the left by $\bra{P_0}$, and from the right by $|1]$, then we obtain
	\begin{equation}
	\begin{aligned}
	\frac{\ip{nP_0}}{\ip{P_0i+1}}=\frac{[i+11]}{[n1]}
	\end{aligned}
	\end{equation}
	Now we express $\frac{\ip{i+1n}}{\ip{P_0i+1}}$ multiplying~(\ref{eq:MC1}) from the left by $\bra{i+1}$, and from the right by $|1]$ which yields $\ip{i+1n}=\ip{P_0i+1}$ since $[P_0|=-[n|$. 
	The delta function $\delta(P_0^2)$, given that $\de{4}{P_0+P_{i+1}+P_n}$, equals
	\begin{equation}
	\delta((P_{i+1}+P_n)^2)=\delta(2P_{i+1}\cdot P_n)=\delta(\ip{i+1n}[ni+1])=\frac{\delta([ni+1])}{\ip{i+1n}}
	\end{equation}
	Total numerical multiplier under the integral sign~(\ref{eq:int1}) takes the form 
	\begin{equation}
	\frac{\ip{P_0i+1}^3}{\ip{i+1n}^2\ip{nP_0}}=\frac{\ip{i+1n}}{\ip{nP_0}},
	\end{equation}
	where $\ip{i+1n}=\ip{P_0i+1}$ is used. Let us calculate it, multiplying~(\ref{eq:MC1}) from the left by $\bra{n}$, we get
	\begin{equation}
	-\ip{nP_0}[P_0|=\ip{ni+1}[i+1|=\ip{ni+1}\frac{[i+11]}{[n1]}[n|
	\end{equation}
	Since $-[P_0|=[n|$, then
	\begin{equation}
	\frac{\ip{i+1n}}{\ip{nP_0}}=\frac{[1n]}{[i+11]}
	\end{equation}
	Finally, performing the integration over $P_0$ in the last line of Eq.~(\ref{eq:int1}), we arrive at
	\begin{equation}
	\begin{aligned}
	&\delta((P_{i+1}+P_n)^2)M(1,2...\frac{\ip{i+1n}}{\ip{P_0i+1}}\eta_n,\ket{-P_0},|-P_0])\frac{\ip{P_0i+1}^3}{\ip{i+1n}\ip{nP_0}}\cdot\\
	&\cdot\de{4}{\eta_{i+1}-\frac{\ip{nP_0}}{\ip{P_0i+1}}\eta_n}=\\
	&=\delta([i+1n])\frac{[1n]}{[i+11]}M(1,2...\la_n+\frac{[i+11]}{[n1]}\la_{i+1},\lat_n,\eta_n)\de{4}{\eta_{i+1}-\frac{[i+11]}{[n1]}\eta_n}
	\end{aligned}
	\end{equation}
	The result obtained coincides with the expression~(\ref{eq:int2}), i.e the left hand side of Eq.~(\ref{eq:th_alg}), that completes the proof.
    \clearpage
	\bibliographystyle{JHEP.bst}
	\bibliography{main}
\end{document}